\documentclass[11pt,a4paper]{article}
\usepackage[utf8]{inputenc}
\usepackage{amsmath,amssymb}
\usepackage{xcolor}
\usepackage{graphicx}
\usepackage{tikz}
\usetikzlibrary{arrows.meta}
\usepackage[margin=25mm]{geometry}
\usepackage[numbers,sort&compress]{natbib}
\usepackage[colorlinks=true,linkcolor=blue,citecolor=blue,urlcolor=blue]{hyperref}
\numberwithin{equation}{section}
\allowdisplaybreaks

\newcommand{\ket}[1]{\lvert #1\rangle}
\newcommand{\doi}[1]{\href{https://doi.org/#1}{doi:\nolinkurl{#1}}}
\newcommand{\arxiv}[1]{\href{https://arxiv.org/abs/#1}{arXiv:#1}}
\newcommand{\cK}{\mathcal{K}}
\newcommand{\cL}{\mathcal{L}}
\newcommand{\cP}{\mathcal{P}}

\newcommand{\nh}{\mathfrak{N}}
\newcommand{\sz}{\sigma_z}
\newcommand{\sx}{\sigma_x}
\newcommand{\ad}{a_{+}^{\dagger}}
\newcommand{\am}{a_{-}}

\newcommand{\Hm}{\widetilde{H}}
\newcommand{\sK}{\mathsf{K}}

\begin{document}

  \title{The hidden symmetry operators of the asymmetric quantum Rabi model and its two-photon and two-mode variants in closed form}

  \author{Marcus Kollar and Arka Dutta\\[4pt]
    \normalsize Theoretical Physics III, Center for Electronic Correlations and Magnetism,\\
    \normalsize Institute of Physics, University of Augsburg, 86135 Augsburg, Germany}

  \date{November 9, 2026}

  \maketitle

  \begin{abstract}
    The asymmetric quantum Rabi model, a two-level system with
    splitting $\Delta$ and bias $\epsilon$ coupled with strength $g$
    to a boson mode, has level crossings when $\epsilon$ is an integer
    multiple $N$ of the mode frequency. They are due to a hidden
    symmetry operator $J$, for which no general formula has been
    available. Here we obtain $J$ in closed form for every $N$. We
    find that $J$ is given by the parity of the symmetric model times
    the continuant (ordered tridiagonal determinant) of $N$
    first-order operators, which are the Hamiltonians at the
    intermediate biases $N-1,N-3,\dots,1-N$ minus a common number
    operator. The symmetry is thus a staircase of $N$ bias steps that
    closes on the mirror image of the model precisely at integer
    bias. The proof uses induction on two commutation relations. The
    same relations hold in the asymmetric two-photon and two-mode
    models, written with the generators of $\mathfrak{su}(1,1)$, so
    that the same continuant gives their symmetry operators for every
    order, previously known only for the lowest ones. We also relate
    the closed formula for $J$ to the known recursion relations for
    its coefficients and give their terms of lowest order in $\Delta$
    explicitly.
  \end{abstract}
  
  \clearpage

  \section{Introduction}
  \label{sec:intro}
  
  The quantum Rabi model
  (QRM)~\cite{rabi_process_1936,rabi_space_1937,jaynes_comparison_1963,braak_semiclassical_2016,xie_quantum_2017}
  describes a two-level system coupled linearly to a single bosonic
  mode and is the simplest fully quantum model of light-matter
  interaction. It is central to cavity and circuit quantum
  electrodynamics (QED), trapped ions and related
  platforms~\cite{friskkockum_ultrastrong_2019,forndiaz_ultrastrong_2019,blais_circuit_2021}. Since
  the ultrastrong- and deep-strong-coupling regimes have been reached
  in superconducting
  circuits~\cite{niemczyk_circuit_2010,forndiaz_observation_2010,yoshihara_superconducting_2017,yoshihara_inversion_2018},
  where the coupling is comparable to or larger than the oscillator
  and qubit frequencies, the rotating-wave approximation fails and the
  full QRM, including its counter-rotating terms, is
  needed~\cite{friskkockum_ultrastrong_2019,forndiaz_ultrastrong_2019}. In
  most of these experiments the qubit is a flux qubit, whose two
  persistent-current states are split by a tunnel coupling $\Delta$
  and offset by an energy bias $\epsilon$ that is set by the magnetic
  flux through the qubit loop and tuned in
  situ~\cite{yoshihara_superconducting_2017}. The model with bias is
  the asymmetric quantum Rabi model (AQRM), Eq.~\eqref{eq:HAQRM}
  below. It is the simplest model in which a two-level system couples
  to a bosonic mode without any obvious symmetry, and it is
  nevertheless exactly solvable: Braak's solution in Bargmann
  space~\cite{braak_integrability_2011} was reproduced with Bogoliubov
  operators~\cite{chen_exact_2012}. Special values of the bias make
  the AQRM a test case for the relation between symmetry, level
  crossings and integrability, which is the subject of this paper.
  
  The QRM is integrable~\cite{braak_integrability_2011}. Its
  $\mathbb{Z}_2$ parity $e^{i\pi a^\dagger a}\sigma_z$ splits the
  Hilbert space into two invariant subspaces, and levels of opposite
  parity cross as functions of the coupling~$g$. The bias breaks this
  parity, and generically the crossings disappear. Braak observed,
  however, that they reappear at special values of the
  bias~\cite{braak_integrability_2011}. Li and Batchelor derived
  algebraic equations for the exceptional eigenvalues of the biased
  model and confirmed the crossings
  numerically~\cite{li_algebraic_2015}. Wakayama proved the crossings
  for the first nontrivial bias, using the representation theory of
  $\mathfrak{sl}_2$, and conjectured the general
  case~\cite{wakayama_symmetry_2017}. Kimoto, Reyes-Bustos and
  Wakayama then showed, via determinant expressions for the constraint
  polynomials, that the spectrum is degenerate if and only if the bias
  takes these special values, and that non-Juddian exceptional
  eigenvalues are never degenerate~\cite{kimoto_determinant_2021}. The
  key step is a divisibility relation between the constraint
  polynomials at the two matched baseline energies, whose quotient is
  an explicit tridiagonal determinant $A^\ell_N$ in the squared
  coupling and the squared qubit splitting.
  
  We write the AQRM as
  \begin{align}
    H_{\rm AQRM}=\frac{\Delta}{2}\,\sigma_z+\frac{\epsilon}{2}\,\sigma_x+a^\dagger a
    +g\,(a^\dagger+a)\,\sigma_x ,
    \label{eq:HAQRM}
  \end{align}
  with $\omega=1$, as Xie and Chen~\cite{xie_general_2022}. The hidden
  symmetry then occurs for $\epsilon=N\in\mathbb{Z}$ (by symmetry we
  take $N\ge0$). Other works use conventions that differ by factors of
  two. All statements below are converted to Eq.~\eqref{eq:HAQRM}.
  
  Level crossings without an obvious symmetry suggest a hidden
  one. Ashhab~\cite{ashhab_attempt_2020} examined numerically how the
  eigenstates distribute over putative symmetry classes and found
  that, unlike for an ordinary symmetry, the distribution depends on
  the parameters. Li and Batchelor~\cite{li_hidden_2021} determined
  the conditions under which the hidden symmetry exists in several
  related biased models and linked it to selective tunnelling in the
  displaced-oscillator basis. Mangazeev, Batchelor and
  Bazhanov~\cite{mangazeev_hidden_2021} found operators $J$ that
  commute with the Hamiltonian at the first few special biases, of the
  form of the parity operator $e^{i\pi a^\dagger a}$ times a
  $2\times2$ matrix of polynomials in $a$ and $a^\dagger$.
  Reyes-Bustos, Braak and Wakayama~\cite{reyesbustos_remarks_2021}
  showed that such a $J$ exists for every integer bias and is unique
  up to a factor within its degree, that it is self-adjoint, and that
  $J^2$ is a polynomial $p_N(H)$ of degree $N$ in $H$. Reyes-Bustos
  and Wakayama~\cite{reyesbustos_degeneracy_2022} conjectured that
  $p_N$, evaluated at the Juddian baseline energies, equals the
  quotient $A^N_n$ of constraint polynomials of
  Ref.~\cite{kimoto_determinant_2021}, and verified this for
  $N\le6$. From the conjecture they derived a tridiagonal determinant
  expression for $p_N$ and found that the $N$ lowest energy curves are
  approximated by the curves $p_N(E)=0$ at large coupling. Both groups
  state that no general formula for $J$ is
  known~\cite{mangazeev_hidden_2021,reyesbustos_degeneracy_2022} (see
  also Ref.~\cite{lu_hidden_2022}), and the conjecture has remained
  open.
  
  These constructions have been extended to the biased Dicke
  model~\cite{lu_hidden_2021}, to the anisotropic AQRM and the
  asymmetric Rabi-Stark models~\cite{lu_hidden_2022}, to the
  asymmetric two-mode model for the lowest nontrivial
  bias~\cite{yan_analytical_2022}, and to the asymmetric two-photon
  model. There, crossings in a subspace appear when the bias is an
  even multiple of the renormalized cavity
  frequency~\cite{xie_double_2021}, and the symmetry operators were
  derived hierarchically for the lowest three such
  biases~\cite{xie_symmetry_2022}. For the two-photon model,
  Reyes-Bustos and Wakayama have recently proved, from the monodromy
  of the underlying Fuchsian equation, the existence of the
  degeneracies and of the symmetry operators, and characterized the
  degeneracies completely~\cite{reyesbustos_twophoton_2026}.
  
  For the AQRM itself, Xie and Chen~\cite{xie_general_2022} gave a
  scheme within a Bogoliubov operator approach that yields the
  symmetry operator for any given integer bias. In a rotated qubit
  basis and in terms of the shifted (Bogoliubov) operators
  $a_\pm=a\pm g$, the entries of $J$ are expanded in normal-ordered
  monomials, and the coefficients follow level by level from recursion
  relations (Sec.~\ref{sec:model}). The Bogoliubov frame is also the
  natural one for the physics. At $\Delta=0$ the two displaced
  oscillators are exactly degenerate level by level when $\epsilon=N$,
  which is the elementary reason for the crossings
  (Sec.~\ref{sec:chain}), and it is the frame in which the closed
  formula of this paper takes its simplest form. The physical origin
  of the symmetry was studied in several works. In
  Ref.~\cite{semple_asymptotic_2018} $\Delta$ was treated as the
  perturbation. At integer bias the $N$ lowest states are then
  unpaired, and every higher level of the $\Delta=0$ problem is a
  degenerate pair that splits at first order in $\Delta$. Yang, Chen
  and Luo~\cite{yang_origin_2025} studied the opposite regime
  $\Delta\gg\omega$ and traced the hidden symmetry to level matching
  in the asymmetric double-well potential.
  
  Operators that intertwine two Hamiltonians are the basic objects of
  supersymmetric quantum mechanics and of the Darboux-Crum
  transformations (see Ref.~\cite{rosu_short_1999} for a survey). A
  differential operator of order $N$ that intertwines two
  one-dimensional Schr\"odinger Hamiltonians defines an $N$-fold or
  polynomial
  supersymmetry~\cite{aoyama_nfold_2001,andrianov_nonlinear_2003}. Two
  independent intertwiners between the same Hamiltonians combine to a
  hidden symmetry
  operator~\cite{andrianov_nonlinear_2003,andrianov_hidden_2009}, also
  for $2\times2$ matrix Hamiltonians~\cite{andrianov_matrix_1997}, and
  partner Hamiltonians that differ only by a translation or reflection
  are called self-isospectral~\cite{andrianov_hidden_2009}. For the
  Jaynes-Cummings model, matrix intertwiners connect Hamiltonians at
  different detunings~\cite{ates_susy_2025}. The hidden symmetry of
  the AQRM is of this kind. Its operator intertwines the Hamiltonian
  at bias $N$ with its mirror image at bias $-N$, and the closed
  formula given below builds this intertwiner from $N$ first-order
  steps.
  
  In this paper we give the symmetry operator in closed form for every
  integer bias (Sec.~\ref{sec:formula}). In the frame of
  Ref.~\cite{xie_general_2022} the Hamiltonian at bias $\epsilon$ is,
  up to a constant, the sum of a first-order operator $\cK$, the bias
  term $\frac{\epsilon}{2}\sz$, and the non-Hermitian number operator
  $\nh=\ad\am$ of the shifted operators (Sec.~\ref{sec:model}). The
  symmetry operator is the parity of the symmetric model times the
  continuant of the $N$ operators $\cK-c_m\sz$, $c_m=\frac{N+1}{2}-m$,
  that is, the determinant of the tridiagonal matrix with these
  operators on the diagonal, kept in their order, and with the numbers
  $m(N-m)/4$ as products of the off-diagonal entries. Each factor maps
  eigenstates at one bias to eigenstates at the bias lowered by two,
  up to a constant, and the continuant is the combination of these $N$
  steps in which the constants cancel (Fig.~\ref{fig:staircase}). The
  proof is an induction of a few lines that rests on two commutation
  relations. At $\Delta=0$ the formula reduces to the $N$th powers of
  the two shifted ladder operators, which exchange the members of the
  degenerate pairs of the two displaced oscillators
  (Sec.~\ref{sec:chain}). In Sec.~\ref{sec:su11} we show that the same
  two commutation relations hold in the asymmetric two-photon and
  two-mode models, written in terms of the generators of
  $\mathfrak{su}(1,1)$, so that the same continuant gives their
  symmetry operators for every order. In Sec.~\ref{sec:coefficients}
  we connect the closed formula with the recursion relations of
  Ref.~\cite{xie_general_2022} for the coefficients, which depend on
  $N$ only through their distance from the highest degree, and give
  their lowest orders in $\Delta$.  Sec.~\ref{sec:summary} provides a
  summary and outlook. In particular the polynomial $p_N$ in
  $J^2=p_N(H)$ and the consequences of the closed formula for the
  spectrum will be addressed in a separate publication.
  
  \section{The model and the intertwining condition}
  \label{sec:model}
  
  After the rotation of the qubit by $\pi/2$ about the $y$ axis,
  $e^{i\pi\sigma_y/4}\sigma_x e^{-i\pi\sigma_y/4}=\sigma_z$, and in
  terms of the shifted operators $a_\pm=a\pm g$, the
  Hamiltonian~\eqref{eq:HAQRM} at bias $\epsilon$
  reads~\cite{xie_general_2022}
  \begin{align}
    H_\epsilon=
    \begin{pmatrix} \ad a_{+}-g^2+\frac{\epsilon}{2} & -\frac{\Delta}{2}\\[3pt]
      -\frac{\Delta}{2} & a_{-}^{\dagger}\am-g^2-\frac{\epsilon}{2}\end{pmatrix}.
    \label{eq:Hrot}
  \end{align}
  The symmetry operator at $\epsilon=N$ is written as $J=PQ$ with the
  parity operator $P=e^{i\pi a^\dagger a}$ and a $2\times2$ matrix $Q$
  of operators. Since $PaP=-a$ and hence $Pa_\pm P=-a_\mp$,
  conjugation with $P$ interchanges $\ad a_+$ and $a_-^\dagger\am$ in
  Eq.~\eqref{eq:Hrot}, and we write $\Hm_\epsilon=PH_\epsilon P$. The
  condition $[J,H_N]=0$ is then equivalent to~\cite{xie_general_2022}
  \begin{align}
    Q\,H_N=\Hm_N\,Q .
    \label{eq:QH}
  \end{align}
  Here Ref.~\cite{xie_general_2022} expands the entries of
  $Q=\bigl(\begin{smallmatrix}A&B\\C&D\end{smallmatrix}\bigr)$ in
  normally ordered monomials of total degree at most $N$ (the minimal
  degree, as the form of $J$ is not
  unique~\cite{reyesbustos_remarks_2021}),
  \begin{align}
    X=\sum_{i,j\ge0,\ i+j\le N}X_{i,j}\,(\ad)^i(\am)^j,\qquad X=A,B,C,D,
    \label{eq:expansion}
  \end{align}
  where $[\am,\ad]=1$, so that $\ad$ and $\am$ generate the Weyl
  algebra, with $a_+=\am+2g$ and $a_-^\dagger=\ad-2g$. With the
  normalization $B_{0,N}=1$ the coefficients are then determined by
  recursion relations (Sec.~\ref{sec:coefficients}).
  
  The closed formula is best stated in a slightly different form. The
  operator
  \begin{align}
    \nh=\ad\am
  \end{align}
  is a multiple of the unit matrix. It is the number operator of the
  pseudo-bosonic pair $\am$, $\ad$ in the sense of
  Bagarello~\cite{bagarello_pseudobosons_2010,bagarello_examples_2010}. It
  is not Hermitian, $\nh^\dagger=a_-^\dagger a_+$, but it obeys
  $[\nh,\ad]=\ad$ and $[\nh,\am]=-\am$, and it is similar to the
  number operator,
  $\nh=e^{g(a+a^\dagger)}\,a^\dagger a\,e^{-g(a+a^\dagger)}$. Only the
  commutation relations will be used. Since $\ad a_+=\nh+2g\ad$ and
  $a_-^\dagger\am=\nh-2g\am$,
  \begin{align}
    H_\epsilon+g^2=\cK+\nh+\frac{\epsilon}{2}\,\sz,\qquad
    \cK=\begin{pmatrix}2g\,\ad & -\frac{\Delta}{2}\\[2pt] -\frac{\Delta}{2} & -2g\,\am\end{pmatrix}.
    \label{eq:HK}
  \end{align}
  The operator $\cK$ is of first order in $\ad$, $\am$. All that will
  be used about $\cK$, $\nh$ and $\sz$ are the relations
  \begin{align}
    [\cK,\nh]=-\tfrac12\bigl(\sz\cK+\cK\sz\bigr),\qquad [\sz,\nh]=0,\qquad \sz^2=1 .
    \label{eq:relations}
  \end{align}
  The first one holds because $[\ad,\nh]=-\ad$ and $[\am,\nh]=\am$, so
  that the commutator of the diagonal part of $\cK$ with $\nh$ is
  $-\sz$ times that part, while the off-diagonal part of $\cK$
  commutes with $\nh$ and anticommutes with $\sz$. Finally,
  $\sx\Hm_N\sx=H_{-N}$, because interchanging the two oscillators and
  then the two rows and columns amounts to reversing the sign of the
  bias. Multiplying Eq.~\eqref{eq:QH} by $\sx$ from the left, the
  intertwining condition becomes
  \begin{align}
    \cP\,H_N=H_{-N}\,\cP,\qquad \cP=\sx Q ,
    \label{eq:intertwine}
  \end{align}
  where $g^2$ may be added to both Hamiltonians. The symmetry operator maps the model at bias $N$ to the
  model at bias $-N$, which is unitarily equivalent to it. The equivalence is implemented by
  \begin{align}
    \Pi=P\sx ,\qquad \Pi H_\epsilon\Pi=H_{-\epsilon},\qquad \Pi\cK\Pi=\cK^\dagger,\qquad \Pi\sz\Pi=-\sz ,
    \label{eq:Pi}
  \end{align}
  the parity operator of the symmetric model, where $\Pi\ad\Pi=-a_-^\dagger$ and $\Pi\am\Pi=-a_+$ were
  used. In terms of $\cP$, the symmetry operator is $J=PQ=\Pi\cP$.
  
  \section{The symmetry operator in closed form}
  \label{sec:formula}
  
  \subsection{Statement}
  
  For a positive integer $N$ let
  \begin{align}
    \cL_m=\cK-c_m\,\sz,\qquad c_m=\frac{N+1}{2}-m,\qquad m=1,\dots,N ,
    \label{eq:Lm}
  \end{align}
  and define operators $\cP_0,\dots,\cP_N$ by the three-term recurrence
  \begin{align}
    \cP_0=1,\qquad \cP_1=\cL_1,\qquad
    \cP_m=\cP_{m-1}\,\cL_m+d_{m-1}\,\cP_{m-2}\quad(m\ge2),\qquad
    d_m=\frac{m(N-m)}{4} .
    \label{eq:formula}
  \end{align}
  Then $\cP_N$ satisfies Eq.~\eqref{eq:intertwine}, and
  \begin{align}
    Q=\begin{pmatrix}A&B\\C&D\end{pmatrix}=\Bigl(-\frac{1}{2g}\Bigr)^{N}\sx\,\cP_N ,\qquad
    J=e^{i\pi a^\dagger a}\,Q=\Bigl(-\frac{1}{2g}\Bigr)^{N}\Pi\,\cP_N ,
    \label{eq:Q}
  \end{align}
  is the symmetry operator of Ref.~\cite{xie_general_2022} with its
  normalization $B_{0,N}=1$. This is the main result of the paper, and
  it is proved in Sec.~\ref{sec:proof}. For $N=0$,
  Eqs.~\eqref{eq:formula} and \eqref{eq:Q} give $\cP_0=1$, $Q=\sx$ and
  $J=\Pi$, the parity of the symmetric model (cf.\
  Ref.~\cite{reyesbustos_remarks_2021}), with $J^2=1$. The hidden
  symmetry at bias $N$ is thus the continuation of this parity to
  integer bias.
  
  The recurrence \eqref{eq:formula} is that of a continuant, the
  determinant of a tridiagonal matrix (see, e.g.,
  Ref.~\cite[Sec.~6.7]{graham_concrete_1994} and
  Ref.~\cite[Chap.~XIII]{muir_treatise_1960}). For numbers
  $x_1,\dots,x_n$ on the diagonal, $\xi_m$ above and $\eta_m$ below
  it, the determinant obeys
  $K_n=x_nK_{n-1}-\xi_{n-1}\eta_{n-1}K_{n-2}$.
  Equation~\eqref{eq:formula} is this recurrence with the operators
  $\cL_m$ as diagonal entries, with $-\xi_m\eta_m=d_m$, and with the
  convention that the products are ordered. Expanding the recurrence
  as in Euler's rule for continuants, $\cP_N$ is the sum over all ways
  of removing disjoint pairs of neighbours from the ordered product
  $\cL_1\cL_2\cdots\cL_N$ and replacing each removed pair
  $\cL_m\cL_{m+1}$ by the weight $d_m$,
  \begin{align}
    \cP_N={}&\cL_1\cdots\cL_N+\sum_{m=1}^{N-1}d_m\,\cL_1\cdots\cL_{m-1}\,\cL_{m+2}\cdots\cL_N
              \nonumber\\
            &+\sum_{m+1<m'}d_m\,d_{m'}\,\cL_1\cdots\widehat{\cL_m\cL_{m+1}}\cdots
              \widehat{\cL_{m'}\cL_{m'+1}}\cdots\cL_N+\dots ,
              \label{eq:euler}
  \end{align}
  where the hats mark the removed pairs. No two operators are ever
  interchanged, so the noncommutativity of the $\cL_m$ causes no
  ambiguity. The same sum is obtained by building the recurrence from
  the other end, so that
  $\cP_N=\cL_1\cP'_{N-1}+\frac{N-1}{4}\cP'_{N-2}$ with the continuants
  $\cP'_k$ of $\cL_{N-k+1},\dots,\cL_N$.
  
  The numbers in Eq.~\eqref{eq:formula} are those of a spin
  $s=(N-1)/2$. The $c_m$ are the eigenvalues of $S_z$, listed in
  decreasing order, and $4d_m=m(N-m)=(s+c_m)(s-c_m+1)$ is the product
  of the matrix elements
  $\langle c_m|S_+|c_m-1\rangle=\langle
  c_m-1|S_-|c_m\rangle=\sqrt{(s+c_m)(s-c_m+1)}$ of $S_+$ and $S_-$
  between the neighboring states $c_m$ and $c_m-1$. Formally,
  therefore,
  \begin{align}
    \cP_N=\det\nolimits_{s=(N-1)/2}\Bigl[\,\cK-\sz S_z+\tfrac12\,(S_+-S_-)\Bigr],
    \label{eq:det}
  \end{align}
  with the determinant over the spin space understood as the ordered
  continuant. Since $\cL_m=H_{-2c_m}+g^2-\nh$, the $N$ factors are the
  Hamiltonians at the intermediate biases $-2c_m=1-N,3-N,\dots,N-1$
  minus $\nh$. The meaning of this chain is discussed in
  Sec.~\ref{sec:chain}.
  
  The first cases are, with
  \begin{align}
    K=\frac{\cK}{2g}=\begin{pmatrix}\ad & -w\\ -w & -\am\end{pmatrix},\qquad w=\frac{\Delta}{4g},\qquad
    W=[K,\sz]=2w\begin{pmatrix}0&1\\-1&0\end{pmatrix},
    \label{eq:K}
  \end{align}
  and $P_N=\cP_N/(2g)^N$,
  \begin{align}
    P_1=K,\qquad
    P_2=K^2+\frac{W}{4g},\qquad
    P_3=K^3+\frac{KW+WK}{2g}+\frac{[W,\sz]}{8g^2} ,
    \label{eq:P123}
  \end{align}
  so that $Q=(-1)^N\sx P_N$ gives
  \begin{align}
    Q^{(1)}=
    \begin{pmatrix}w&\am\\-\ad&w\end{pmatrix},\qquad
    Q^{(2)}=
    \begin{pmatrix} w(\am-\ad)-\dfrac{w}{2g} & \am^2+w^2\\[8pt]
      \ad{}^2+w^2 & w(\am-\ad)+\dfrac{w}{2g}\end{pmatrix},
    \label{eq:Q12}
  \end{align}
  in agreement with Ref.~\cite{xie_general_2022} and, for $N=1,2$,
  with Ref.~\cite{mangazeev_hidden_2021}. For $N=1$ the symmetry
  operator is simply $J=-\Pi K$. The operators for $N\le20$ are
  available as a data set~\cite{kollar_symmetry_2026}
  (Sec.~\ref{sec:data}). Dividing the recurrence~\eqref{eq:formula} by
  $(2g)^m$ gives
  $P_m=P_{m-1}\bigl(K-\frac{c_m}{2g}\sz\bigr)+\frac{d_{m-1}}{4g^2}P_{m-2}$. Since
  $K$ contains $\Delta$ and $g$ only through $w$, the operator $Q$ is
  a polynomial in $1/g$ at fixed $w$, and
  \begin{align}
    \lim_{g\to\infty,\ w\ \text{fixed}}Q=\sx\,(-K)^N .
    \label{eq:limit}
  \end{align}
  At strong coupling and fixed $\Delta/g$ the symmetry operator is
  thus the parity times the $N$th power of a single first-order
  operator.

  \subsection{Proof}
  \label{sec:proof}

  Let $\mathcal{H}_m\equiv H_{2m-N}+g^2$, so that $\mathcal{H}_N$ and
  $\mathcal{H}_0$ are the Hamiltonians at the biases $N$ and $-N$,
  plus $g^2$. A single factor $\cL_m$ maps the Hamiltonian at bias
  $2m-N$ to that at bias $2m-N-2$ up to a constant, since by
  Eq.~\eqref{eq:relations}
  \begin{align}
    \cL_m\mathcal{H}_m-\mathcal{H}_{m-1}\cL_m
    &=[\cK,\nh]+\Bigl(m-\frac N2\Bigr)\cK\sz-\Bigl(m-1-\frac N2\Bigr)\sz\cK-c_m[\sz,\cK]-c_m
      \nonumber\\
    &=\Bigl(m-\frac{N+1}{2}+c_m\Bigr)\bigl(\cK\sz-\sz\cK\bigr)-c_m=-c_m .
      \label{eq:onestep}
  \end{align}
  The leftover constants are removed by the second term of the recurrence. We show by induction that
  \begin{align}
    \cP_m\mathcal{H}_m-\mathcal{H}_0\,\cP_m=-2d_m\,\cP_{m-1},\qquad m=0,1,\dots,N .
    \label{eq:quasi}
  \end{align}
  For $m=0$ both sides vanish, and for $m=1$ this is
  Eq.~\eqref{eq:onestep} with $c_1=(N-1)/2=2d_1$. For $m\ge2$ we
  insert Eq.~\eqref{eq:formula}, use Eq.~\eqref{eq:onestep},
  $\mathcal{H}_m=\mathcal{H}_{m-2}+2\sz$, and Eq.~\eqref{eq:quasi} for
  $m-1$ and $m-2$:
  \begin{align}
    \cP_m\mathcal{H}_m
    ={}&\cP_{m-1}\bigl(\mathcal{H}_{m-1}\cL_m-c_m\bigr)+d_{m-1}\cP_{m-2}\bigl(\mathcal{H}_{m-2}+2\sz\bigr)
         \nonumber\\
    ={}&\bigl(\mathcal{H}_0\cP_{m-1}-2d_{m-1}\cP_{m-2}\bigr)\cL_m-c_m\cP_{m-1}
         +d_{m-1}\bigl(\mathcal{H}_0\cP_{m-2}-2d_{m-2}\cP_{m-3}+2\cP_{m-2}\sz\bigr)
         \nonumber\\
    ={}&\mathcal{H}_0\cP_m-c_m\cP_{m-1}-2d_{m-1}\cP_{m-2}\bigl(\cL_m-\sz\bigr)-2d_{m-1}d_{m-2}\,\cP_{m-3}.
  \end{align}
  Since $c_m=c_{m-1}-1$, we have $\cL_m-\sz=\cL_{m-1}$, and by the recurrence
  $\cP_{m-2}\cL_{m-1}=\cP_{m-1}-d_{m-2}\cP_{m-3}$. The terms with $\cP_{m-3}$ cancel, and
  \begin{align}
    \cP_m\mathcal{H}_m=\mathcal{H}_0\cP_m-\bigl(c_m+2d_{m-1}\bigr)\cP_{m-1},\qquad
    c_m+2d_{m-1}=2d_m ,
  \end{align}
  which completes the induction. The right-hand side of
  Eq.~\eqref{eq:quasi} vanishes for $m=N$ (and trivially for $m=0$),
  so that $\cP_N$ satisfies Eq.~\eqref{eq:intertwine} and maps the
  eigenstates of $H_N$ to eigenstates of $H_{-N}$ with the same
  energy. Its leading terms are those of $\cK^N$, namely
  $(2g)^N\mathrm{diag}\bigl(\ad{}^N,(-\am)^N\bigr)$, so that $Q$ in
  Eq.~\eqref{eq:Q} has entries of degree at most $N$ with $B_{0,N}=1$
  and $C_{N,0}=(-1)^N$. The polynomial solution of Eq.~\eqref{eq:QH}
  of degree $N$ is unique up to a factor
  (Ref.~\cite{reyesbustos_remarks_2021}, Proposition~3.1 of the arXiv
  version). Within the expansion~\eqref{eq:expansion} this also
  follows from the recursion of Ref.~\cite{xie_general_2022}, see
  Sec.~\ref{sec:recursion}. Therefore Eq.~\eqref{eq:Q} is the operator
  of Ref.~\cite{xie_general_2022}.

  \subsection{Chain structure}
  \label{sec:chain}

  Nothing in the argument requires $N$ to be an integer. For an
  arbitrary real bias $\nu$ in place of $N$, Eqs.~\eqref{eq:Lm} and
  \eqref{eq:formula} define operators $\cP^{(\nu)}_m$ for all $m\ge0$,
  and the induction gives
  \begin{align}
    \cP^{(\nu)}_m\bigl(H_{2m-\nu}+g^2\bigr)-\bigl(H_{-\nu}+g^2\bigr)\cP^{(\nu)}_m
    =-\frac{m(\nu-m)}{2}\,\cP^{(\nu)}_{m-1} .
    \label{eq:quasinu}
  \end{align}
  The hidden symmetry at integer bias is the statement that this chain
  of operators terminates. The right-hand side vanishes when $m$
  reaches $\nu$, which requires $\nu$ to be a positive integer.  This
  is the operator counterpart of the level matching of the two
  displaced oscillators at $\Delta=0$, which is discussed below.

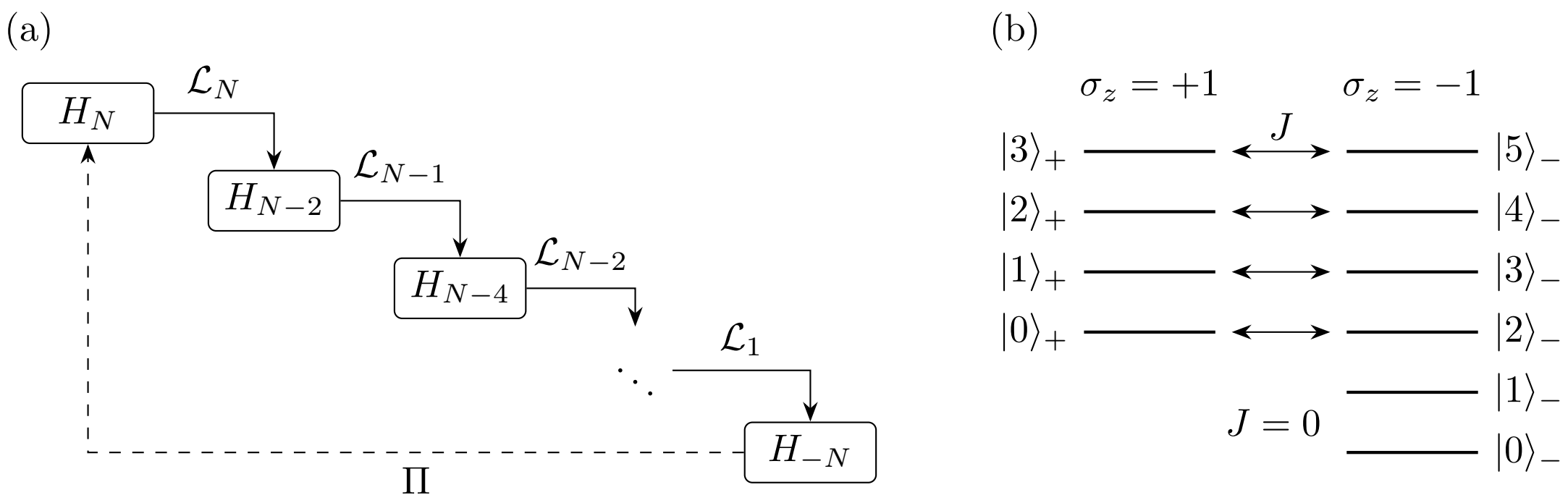
\begin{figure}[t]
    \centering
    \begin{tikzpicture}[>={Stealth[length=1.8mm]},font=\small,
      ham/.style={draw,rounded corners=2pt,inner sep=3pt,minimum width=12mm,minimum height=5.5mm}]
      \node[anchor=west] at (-0.3,3.55) {(a)};
      \node[ham] (h0) at (0.6,2.8) {$H_{N}$};
      \node[ham] (h1) at (2.3,2.0) {$H_{N-2}$};
      \node[ham] (h2) at (4.0,1.2) {$H_{N-4}$};
      \node (dd) at (5.6,0.45) {$\ddots$};
      \node[ham] (hm) at (7.2,-0.3) {$H_{-N}$};
      \draw[->] (h0.east) -| node[pos=0.25,above] {$\cL_N$} (h1.north);
      \draw[->] (h1.east) -| node[pos=0.25,above] {$\cL_{N-1}$} (h2.north);
      \draw[->] (h2.east) -| node[pos=0.25,above] {$\cL_{N-2}$} (dd.north);
      \draw[->] (dd.east) -| node[pos=0.25,above] {$\cL_{1}$} (hm.north);
      \draw[->,dashed] (hm.west) -| node[pos=0.25,below] {$\Pi$} (h0.south);
      \begin{scope}[shift={(9.7,-0.3)}]
        \node[anchor=west] at (-1.0,3.85) {(b)};
        \node at (0.6,3.35) {$\sigma_z=+1$};
        \node at (3.0,3.35) {$\sigma_z=-1$};
        \foreach \m in {0,...,5} {
          \draw[thick] (2.4,0.55*\m) -- (3.6,0.55*\m) node[right] {$\ket{\m}_-$};
        }
        \foreach \n in {0,...,3} {
          \draw[thick] (0,0.55*\n+1.1) -- (1.2,0.55*\n+1.1);
          \node[left] at (0,0.55*\n+1.1) {$\ket{\n}_+$};
          \draw[<->] (1.35,0.55*\n+1.1) -- (2.25,0.55*\n+1.1);
        }
        \node at (1.8,2.98) {$J$};
        \node[left] at (2.3,0.275) {$J=0$};
      \end{scope}
    \end{tikzpicture}
    \caption{(a) The staircase in the bias. The factor $\cL_m$ of the continuant $\cP_N$ maps the
      Hamiltonian at bias $2m-N$ to the Hamiltonian at bias $2m-N-2$ up to a constant,
      Eq.~\eqref{eq:onestep}. In the continuant the constants cancel, so that $\cP_NH_N=H_{-N}\cP_N$, and
      the parity $\Pi$ of the symmetric model maps $H_{-N}$ back to $H_N$. (b) The levels at $\Delta=0$ for
      $N=2$. The ladders of the two displaced oscillators coincide except for the $N$ lowest states of the
      lower one. The operator $J$ exchanges the two members of each pair and annihilates the unpaired
      states.}
    \label{fig:staircase}
  \end{figure}

  The chain~\eqref{eq:quasinu} may be understood as follows
  (Fig.~\ref{fig:staircase}). Raising the bias by two shifts the upper
  displaced ladder up and the lower one down by one quantum each, so
  that the Hamiltonians at biases $\epsilon$ and $\epsilon-2$ have the
  same spectrum up to this relabelling, and at $\Delta=0$ the step
  operator $\cK=\mathrm{diag}(2g\ad,-2g\am)$ is an exact intertwiner
  between them, $\cK(H_\epsilon+g^2)=(H_{\epsilon-2}+g^2)\cK$, because
  $\ad(\ad a_++\frac\epsilon2)=(\ad a_++\frac{\epsilon-2}{2})\ad$ and
  $\am(a_-^\dagger\am-\frac\epsilon2)=(a_-^\dagger\am-\frac{\epsilon-2}{2})\am$. The
  operator $\cP_N$ thus walks down the bias from $N$ to $-N$ in $N$
  steps of two, arriving at $H_{-N}=\Pi H_N\Pi$, which the parity
  $\Pi$ of the symmetric model maps back to $H_N$, and $J$ is this
  parity composed with the staircase. The integer condition is the
  condition that the staircase reaches the mirror image.
  Equation~\eqref{eq:quasinu} connects $H_\nu$ to $H_{\nu-2m}$, and
  $\nu-2m=-\nu$ requires $\nu=m$. For $\Delta\ne0$ the term
  $-\frac\Delta2\sx$ of $\cK$ switches the qubit component and spoils
  each step by the constant $-c_m$ of Eq.~\eqref{eq:onestep}. The
  continuant, with the weights $d_m$ for deleted pairs of steps, is
  the combination in which these constants cancel, and at $\Delta=0$,
  where the steps with $\cK$ alone are exact, it reduces to $\cK^N$,
  as we now show.

  At $\Delta=0$ all operators are diagonal in the qubit space,
  $\cL_m=\mathrm{diag}(2g\ad-c_m,-2g\am+c_m)$, and the two diagonal
  entries of $\cP_N$ are ordinary continuants of commuting
  entries. For commuting entries $x\mp c_m$ and the weights $d_m$ the
  continuant is the determinant $\det[x\mp S_z+\frac12(S_+-S_-)]$ in
  the spin space, as in Eq.~\eqref{eq:det}. The matrices
  $\mp S_z+\frac12(S_+-S_-)$ are nilpotent, since they are ladder
  operators with respect to $S_x$, so that both determinants equal
  $x^N$, and
  \begin{align}
    \cP_N\big|_{\Delta=0}=\cK^N\big|_{\Delta=0},\qquad
    Q\big|_{\Delta=0}=
    \begin{pmatrix}0 & \am^N\\ (-\ad)^N & 0\end{pmatrix}.
    \label{eq:Q0}
  \end{align}
  The symmetry operator at $\Delta=0$ is thus simply the parity times
  the $N$th powers of the two shifted ladder operators. This is the
  case in which the hidden symmetry can be understood by hand
  [Fig.~\ref{fig:staircase}(b)]. The diagonal blocks of $H_N$ are
  displaced oscillators with eigenstates
  $\ket{n}_\pm=D(\mp g)\ket{n}$, where
  $D(\mp g)=e^{\mp g(a^\dagger-a)}$ is the displacement operator, and
  spectra $n-g^2+N/2$ and $m-g^2-N/2$, $n,m=0,1,2,\dots$ The two
  ladders coincide level by level exactly when the bias is an integer,
  with $m=n+N$. Every level except the $N$ lowest is then exactly
  doubly degenerate for all $g$, the pair being $(\ket{m-N}_+,0)$ and
  $(0,\ket{m}_-)$ with $m\ge N$, while the $N$ lowest states
  $(0,\ket{m}_-)$, $m<N$, are
  unpaired~\cite{semple_asymptotic_2018}. The operator $\am^N$ lowers
  the level of the $a_-$ ladder by $N$, and the bias compensates
  exactly. This is the cancellation that Xie and Chen used to identify
  the highest-degree terms of $B$ and
  $C$~\cite{xie_general_2022}. With the phases fixed by
  $\ket{n}_\pm=D(\mp g)\ket{n}$, so that
  $e^{i\pi a^\dagger a}\ket{n}_\pm=(-1)^n\ket{n}_\mp$, the operator
  $J$ exchanges the two members of each pair and annihilates the
  unpaired states: in the basis $\{(\ket{m-N}_+,0),(0,\ket{m}_-)\}$,
  \begin{align}
    J\big|_{\Delta=0}=(-1)^{m-N}\sqrt{\frac{m!}{(m-N)!}}
    \begin{pmatrix}0&1\\1&0\end{pmatrix},
    \label{eq:Jpair}
  \end{align}
  so that the eigenvectors of $J$ are the symmetric and antisymmetric
  combinations of the two members, with eigenvalues
  $\pm\sqrt{m!/(m-N)!}$, and
  \begin{align}
    J^2\big|_{\Delta=0}=\prod_{k=1}^{N}\Bigl(H_N+g^2-\frac N2+k\Bigr).
    \label{eq:J20}
  \end{align}
  The roots of this polynomial are the energies of the unpaired
  states, which span the kernel of $J$ at $\Delta=0$.

  \subsection{Self-adjointness and the eigenvalues of $J$}
  \label{sec:selfadjoint}

  By Eq.~\eqref{eq:Pi},
  $\Pi\cL_m\Pi=\cK^\dagger+c_m\sz=\cL_{N+1-m}^\dagger$. Conjugating
  the expansion~\eqref{eq:euler} with $\Pi$ therefore reverses the
  order of the factors and takes their adjoints, while the weights
  satisfy $d_m=d_{N-m}$. Hence $\Pi\cP_N\Pi=\cP_N^\dagger$, and
  \begin{align}
    J^\dagger=J,\qquad J^2=Q^\dagger Q=(2g)^{-2N}\,\cP_N^\dagger\cP_N .
    \label{eq:J2def}
  \end{align}
  In Ref.~\cite{xie_general_2022} the self-adjointness of $J$ is used
  to relate $C$ to $B$. Here it is a consequence of the formula. The
  adjoint is taken here term by term in $a$ and $a^\dagger$, as in
  Ref.~\cite{reyesbustos_remarks_2021}, so that $J$ is a symmetric
  operator on the domain of $(a^\dagger a)^{N/2}$.  The eigenstates of
  $H_N$ can thus be chosen as eigenstates of $J$. By
  Ref.~\cite{reyesbustos_remarks_2021}, $J^2=p_N(H_N)$ with a
  polynomial $p_N$ of degree $N$, for which a determinant expression
  was conjectured in Ref.~\cite{reyesbustos_degeneracy_2022}. The
  eigenvalue of $J$ on an eigenstate of energy $E$ is therefore
  $+\sqrt{p_N(E)}$ or $-\sqrt{p_N(E)}$. The two signs label the two
  symmetry sectors, and the two states of a degenerate level carry
  opposite signs~\cite{reyesbustos_degeneracy_2022}. For $N=0$ the
  staircase is empty and $J=\Pi$ with the eigenvalues $\pm1$. For
  $N\ge1$ these are replaced by an energy-dependent magnitude, so that
  $J$ acts on the spectrum as a dressed parity, and the generalized
  parity $J/\sqrt{p_N(H)}$ of Ref.~\cite{xie_general_2022} is defined
  on all eigenstates with $p_N(E)\ne0$. We do not determine $p_N$ in
  this paper. At $\Delta=0$ it is given by Eq.~\eqref{eq:J20}, and for
  $N=1$ one finds directly from Eq.~\eqref{eq:Q12} that
  \begin{align}
    J^2=H_1+g^2+\tfrac12+w^2,\qquad w=\frac{\Delta}{4g} .
    \label{eq:J2N1}
  \end{align}
  In this case $J$ has no kernel in the Hilbert space for
  $\Delta\ne0$, because $Q\psi=0$ requires that $(\nh+1+w^2)\psi_1=0$,
  whose solution $\psi_1\propto e^{gz}(z+g)^{-1-w^2}$ in the Bargmann
  representation $a^\dagger=z$, $a=d/dz$ is not entire.

  \section{Two-photon and two-mode models}
  \label{sec:su11}

  The proof of Sec.~\ref{sec:proof} uses only the
  decomposition~\eqref{eq:HK} of the Hamiltonian and the
  relations~\eqref{eq:relations}, and the step from the intertwiner to
  the symmetry operator uses an involution with the
  properties~\eqref{eq:Pi}. All of these are also available in the
  asymmetric two-photon and two-mode quantum Rabi models, which are
  usually written as~\cite{xie_symmetry_2022,yan_analytical_2022}
  \begin{align}
    H_{\rm tp}&=\frac{\Delta}{2}\,\sigma_z+\frac{\epsilon}{2}\,\sigma_x+a^\dagger a
                +g\,\bigl(a^{\dagger2}+a^2\bigr)\,\sigma_x ,
                \label{eq:Htp}\\
    H_{\rm tm}&=\frac{\Delta}{2}\,\sigma_z+\frac{\epsilon}{2}\,\sigma_x+a_1^\dagger a_1+a_2^\dagger a_2
                +g\,\bigl(a_1^\dagger a_2^\dagger+a_1a_2\bigr)\,\sigma_x ,
                \label{eq:Htm}
  \end{align}
  with $\omega=1$, where $|g|<\frac12$ in the two-photon and $|g|<1$
  in the two-mode model. After the rotation of the qubit of
  Sec.~\ref{sec:model}, which replaces $\sigma_x$ by $\sz$ and
  $\sigma_z$ by $-\sx$, both take the form
  \begin{align}
    H_\epsilon=2\sK_0+2\kappa\,(\sK_++\sK_-)\,\sz+\frac{\epsilon}{2}\,\sz-\frac{\Delta}{2}\,\sx ,
    \qquad |\kappa|<\tfrac12 ,
    \label{eq:Hsu11}
  \end{align}
  up to a constant, with generators $\sK_0$, $\sK_\pm$ of
  $\mathfrak{su}(1,1)$ that obey $[\sK_0,\sK_\pm]=\pm\sK_\pm$ and
  $[\sK_+,\sK_-]=-2\sK_0$. For the two-photon model,
  $\sK_0=\frac12(a^\dagger a+\frac12)$, $\sK_+=\frac12a^{\dagger2}$,
  $\sK_-=\frac12a^2$ and $\kappa=g$, and $H_\epsilon$ is the rotated
  $H_{\rm tp}$ plus $\frac12$. For the two-mode model,
  $\sK_0=\frac12(a_1^\dagger a_1+a_2^\dagger a_2+1)$,
  $\sK_+=a_1^\dagger a_2^\dagger$, $\sK_-=a_1a_2$ and $\kappa=g/2$,
  and $H_\epsilon$ is the rotated $H_{\rm tm}$ plus $1$. Let
  $\Omega=\sqrt{1-4\kappa^2}$ be the renormalized frequency (denoted
  by $\beta$ in Refs.~\cite{xie_symmetry_2022,yan_analytical_2022})
  and
  \begin{align}
    \nh_\kappa&=\Omega\,\sK_0-\kappa\,(\sK_+-\sK_-),\nonumber\\
    \cK_\kappa&=\frac{\kappa}{\Omega}\Bigl[(\sK_++\sK_-)\,\sz+\Omega\,(\sK_+-\sK_-)+4\kappa\,\sK_0\Bigr]
                -\frac{\Delta}{4\Omega}\,\sx .
                \label{eq:Ksu11}
  \end{align}
  Then
  \begin{align}
    \frac{H_\epsilon}{2\Omega}=\cK_\kappa+\nh_\kappa+\frac{\epsilon}{4\Omega}\,\sz,\qquad
    [\cK_\kappa,\nh_\kappa]=-\tfrac12\bigl(\sz\cK_\kappa+\cK_\kappa\sz\bigr),\qquad[\sz,\nh_\kappa]=0 .
    \label{eq:relsu11}
  \end{align}
  The first equation holds because $\Omega^2+4\kappa^2=1$, and the
  second follows from
  $[\sK_++\sK_-,\nh_\kappa]=-\Omega(\sK_+-\sK_-)-4\kappa\sK_0$,
  $[\sK_+-\sK_-,\nh_\kappa]=-\Omega(\sK_++\sK_-)$ and
  $[\sK_0,\nh_\kappa]=-\kappa(\sK_++\sK_-)$. These are
  Eqs.~\eqref{eq:HK} and \eqref{eq:relations} with
  $\epsilon/(2\Omega)$ in place of the bias. The induction of
  Sec.~\ref{sec:proof} therefore applies without change. The
  continuant $\cP_N[\cK_\kappa]$ of the operators $\cK_\kappa-c_m\sz$
  with the weights $d_m$, Eq.~\eqref{eq:formula}, satisfies
  \begin{align}
    \cP_N[\cK_\kappa]\,H_{2\Omega N}=H_{-2\Omega N}\,\cP_N[\cK_\kappa]
    \label{eq:intsu11}
  \end{align}
  for every positive integer $N$. The mirror image is reached with $\Pi_\kappa=(-1)^{\sK_0-k}\sx$, where
  $k$ is the Bargmann index of the representation, in which $\sK_0$ has the eigenvalues
  $k,k+1,k+2,\dots$ Since $(-1)^{\sK_0-k}$ anticommutes with $\sK_\pm$, one has
  $\Pi_\kappa H_\epsilon\Pi_\kappa=H_{-\epsilon}$, $\Pi_\kappa\cK_\kappa\Pi_\kappa=\cK_\kappa^\dagger$ and
  $\Pi_\kappa\sz\Pi_\kappa=-\sz$ as in Eq.~\eqref{eq:Pi}, so that
  \begin{align}
    J=\Bigl(-\frac{1}{2\kappa}\Bigr)^{N}\,\Pi_\kappa\,\cP_N[\cK_\kappa]
    \label{eq:Jsu11}
  \end{align}
  commutes with $H_{2\Omega N}$ in every representation of the
  discrete series, and $J^\dagger=J$ follows as in
  Sec.~\ref{sec:selfadjoint}. The hidden symmetry of these models thus
  occurs when the bias is an even multiple $2\Omega N$ of the
  renormalized frequency, as found in
  Refs.~\cite{xie_double_2021,yan_analytical_2022}, and its operator
  is given by the same continuant for every $N$.

  For the two-photon model the structure parallels that of
  Sec.~\ref{sec:model}, with squeezed instead of displaced
  operators. With $b_\pm=a\cosh\theta\pm a^\dagger\sinh\theta$ and
  $\tanh2\theta=2g$, which are the Bogoliubov operators of
  Ref.~\cite{xie_symmetry_2022}, one finds
  $[b_-,b_+^\dagger]=1/\Omega$ and
  \begin{align}
    \nh_\kappa=\frac{\Omega}{2}\,b_+^\dagger b_-+\frac14,\qquad
    \cK_\kappa=
    \begin{pmatrix} g\,b_+^{\dagger2} & -\frac{\Delta}{4\Omega}\\[3pt]
      -\frac{\Delta}{4\Omega} & -g\,b_-^{2}\end{pmatrix},
    \label{eq:Ktwophoton}
  \end{align}
  to be compared with Eq.~\eqref{eq:HK}. For $N=1,2,3$ the operator
  $Q=(-\frac{1}{2g})^N\sx\,\cP_N[\cK_\kappa]$ reproduces the operators
  given in Eqs.~(25), (27) and~(29) of
  Ref.~\cite{xie_symmetry_2022}. The factor $e^{i\pi a^\dagger a/2}$
  used there equals $(-1)^{\sK_0-k}$ in the even subspace
  ($k=\frac14$) and $i(-1)^{\sK_0-k}$ in the odd subspace
  ($k=\frac34$). For the two-mode model, where
  $k=\frac12(|a_1^\dagger a_1-a_2^\dagger a_2|+1)$, the factor
  $(-1)^{\sK_0-k}$ differs from the factor $e^{i\pi a_2^\dagger a_2}$
  of Ref.~\cite{yan_analytical_2022} by a sign that depends only on
  the conserved difference $a_1^\dagger a_1-a_2^\dagger a_2$, and for
  $N=1$ Eq.~\eqref{eq:Jsu11} reproduces Eq.~(49) of that reference.

  Since $\cK_\kappa$ is of second order in the boson operators, one
  expects the square of $J$ to be a polynomial of degree $2N$ rather
  than $N$ in the Hamiltonian, as was found for $N\le3$ in
  Ref.~\cite{xie_symmetry_2022}. For $N=1$ we find, with
  $h=H_{2\Omega}/(2\Omega)$,
  \begin{align}
    J^2=h^2-\bigl(k-\tfrac12\bigr)^2+\frac{\Delta^2}{64\kappa^2},
    \label{eq:J2su11}
  \end{align}
  which contains Eq.~(31) of Ref.~\cite{xie_symmetry_2022} and
  Eq.~(50) of Ref.~\cite{yan_analytical_2022}. For the two-photon
  model a conjecture on this polynomial, analogous to that of
  Ref.~\cite{reyesbustos_degeneracy_2022} for the AQRM, has been
  formulated in Ref.~\cite{reyesbustos_twophoton_2026}.

  \section{The coefficients}
  \label{sec:coefficients}

  \subsection{The recursion of Xie and Chen in operator form}
  \label{sec:recursion}

  The closed formula of Sec.~\ref{sec:formula} was found from the
  explicit coefficients of the expansion~\eqref{eq:expansion}, which
  Ref.~\cite{xie_general_2022} determines one level $i+j$ at a time by
  recursion relations. Here we connect the two descriptions. Writing
  out the four entries of Eq.~\eqref{eq:QH}, with
  $h_\pm=a_\pm^\dagger a_\pm$, gives
  \begin{align}
    Ah_+-h_-A&=\tfrac{\Delta}{2}\,(B-C), &
                                           h_+D-Dh_-&=\tfrac{\Delta}{2}\,(B-C),
                                                      \nonumber\\
    [B,h_-]-NB&=\tfrac{\Delta}{2}\,(A-D), &
                                            [h_+,C]-NC&=\tfrac{\Delta}{2}\,(A-D) .
                                                        \label{eq:s1234}
  \end{align}
  Self-adjointness of $J$ (Sec.~\ref{sec:selfadjoint}) is the
  statement $Q^\dagger=PQP$, i.e.\ $A=\sigma(A)$, $D=\sigma(D)$ and
  $C=\sigma(B)$ with the involution $\sigma(X)=PX^\dagger P$, which
  acts on the coefficients as $X_{i,j}\to(-1)^{i+j}X_{j,i}$. The third
  equation gives $A$ in terms of $B$ and $D$, and eliminating $A$ from
  the first with the help of the second, the system reduces to two
  equations for $B$ and $D$,
  \begin{align}
    Dh_--h_+D&=-\tfrac{\Delta}{2}\,\bigl(B-\sigma(B)\bigr),\nonumber\\
    (\mathcal{L}B)\,h_+-h_-(\mathcal{L}B)&=\tfrac{\Delta}{2}\,[h_++h_-,D],\qquad
                                           \mathcal{L}B=[B,h_-]-NB ,
                                           \label{eq:rec}
  \end{align}
  which are Eqs.~(20) and (21) of Ref.~\cite{xie_general_2022} in
  operator form. We represent a normally ordered operator
  $\sum_{i,j}X_{i,j}\,\ad{}^i\am^j$ by its symbol
  $\sum_{i,j}X_{i,j}\,x^iy^j$.  Multiplication by $\ad$ from the left
  and by $\am$ from the right act on the symbol as $x$ and $y$, and
  multiplication by $\am$ from the left and by $\ad$ from the right as
  $y+\partial_x$ and $x+\partial_y$. With
  $\delta=x\partial_x-y\partial_y$ and $\nabla=\partial_x+\partial_y$,
  Eq.~\eqref{eq:rec} then reads
  \begin{align}
    \bigl(\delta+2g(x+y)\bigr)D&=\tfrac{\Delta}{2}\bigl(B-\sigma(B)\bigr),\nonumber\\
    \bigl(-\delta+2g(x+y)+2g\nabla\bigr)\bigl(-\delta-N+2g\partial_x\bigr)B&=\Delta\,(\delta-g\nabla)\,D .
                                                                             \label{eq:recsym}
  \end{align}
  Only the multiplication by $x+y$ raises the level $i+j$ of a
  monomial by one, so that the equations at level $L+1$ express the
  coefficients at level $L$ through those at the levels $L+1$ and
  $L+2$. The first equation determines $D_{i,j}$ at level $L$ from
  $D_{i-1,j+1}$ with the pivot $2g$, and the second determines
  $B_{i,j}$ from $B_{i-1,j+1}$ with the pivot $2g(N+i-j)=2g(k+2i)$,
  where $k=N-i-j$ is the depth of the coefficient below the top
  level. The pivots vanish only at the top level for $(i,j)=(0,N)$,
  which fixes $B_{0,N}$ as the normalization, and $D$ vanishes at the
  top level. Hence the solution of the recursion is unique, and so is
  the polynomial solution of Eq.~\eqref{eq:QH} of degree $N$ in the
  form~\eqref{eq:expansion}. This is the uniqueness used in
  Sec.~\ref{sec:proof}. Each level contains one more equation than
  unknowns, so that the recursion is overdetermined. The closed
  formula provides a solution for every $N$ and thereby shows that it
  is consistent.

  \subsection{Depth and polynomial structure}
  \label{sec:depth}

  The closed forms of Ref.~\cite{xie_general_2022} for the top levels,
  its Eqs.~(23) to (33), contain $N$ only in combinations such as
  $N-i-1$. This is a general feature. At a fixed distance $k=N-i-j$
  from the top level, the coefficients do not depend on $N$ otherwise,
  and
  \begin{align}
    B_{i,j}=(-1)^i f_k(i,j),\qquad D_{i,j}=(-1)^i h_k(i,j),\qquad k=N-i-j ,
    \label{eq:depth}
  \end{align}
  with functions $f_k$, $h_k$ on the quadrant $i,j\ge0$ that are the
  same for all $N$. In terms of Eq.~\eqref{eq:recsym} this holds
  because $N=i+j+k$ removes every explicit $N$ when the equations are
  evaluated at a point of depth $k$. One has $f_0=\delta_{i,0}$ and
  $h_0=0$. For $1\le k\le14$ we have solved the recursion in this form
  with symbolic $i$ and $j$~\cite{kollar_symmetry_2026}, and the
  functions $f_k$ and $h_k$ are then polynomials in $i$ and $j$, with
  coefficients that are polynomials in $\Delta$ and $1/g$. In the
  recursion this is the statement that the division by the pivot
  $2g(k+2i)$ is exact, which is not obvious there. We expect it to
  hold for every $k$. The first depth functions are
  \begin{align}
    f_1&=0,\qquad h_1=\frac{\Delta}{4g},\qquad
         f_2=\frac{\Delta^{2}}{16g^{2}}\,(j+1),\qquad
         h_2=\frac{\Delta}{8g^{2}}\,(i+1)(j+1), \nonumber\\
    f_3&=0,\qquad
         h_3=\Bigl[\frac{\Delta}{32g^{3}}\,(i+2)(j+2)+\frac{\Delta^{3}}{64g^{3}}\Bigr](i+1)(j+1),
         \label{eq:closedforms}\\
    f_4&=\frac{\Delta^{2}}{384g^{4}}\,(i+1)(i+3)(j+1)(j+2)(j+3)
         +\Bigl(\frac{\Delta^{4}}{512g^{4}}-\frac{\Delta^{2}}{32g^{2}}\Bigr)(i+1)(j+1)(j+2),
         \nonumber\\
    h_4&=\Bigl[\frac{\Delta}{192g^{4}}\,(i+3)(j+3)+\frac{\Delta^{3}}{256g^{4}}\Bigr]
         (i+1)(i+2)(j+1)(j+2),\nonumber
  \end{align}
  which are Eqs.~(23) to (33) of Ref.~\cite{xie_general_2022} with
  $N=i+j+k$ inserted. The function $f_5$ does not vanish, e.g.\
  $B_{0,0}=\Delta^2/(16g^3)$ for $N=5$. In this form the symmetry
  $h_k(i,j)=h_k(j,i)$, i.e.\ $D=\sigma(D)$, is visible. The functions
  $f_k$ and $h_k$ for $k\le14$ give every operator with $N\le14$ and
  the top $15$ levels of every operator. The functions $f_k$ and $h_k$
  are even and odd in $\Delta$, respectively, and $f_k=O(\Delta^2)$
  for $k\ge1$.

  \subsection{The expansion in $\Delta$}
  \label{sec:deltaexp}

  The parts of the operators linear in $\Delta$ follow from
  Eq.~\eqref{eq:recsym} in closed form. We use generating functions
  $\sum_Nt^N(\cdots)$ over the bias, with $N\to t\partial_t$ where
  needed, and the abbreviations $p=1/(1+xt)$ and $q=1/(1-yt)$. For the
  linear part of $D$ the source in Eq.~\eqref{eq:recsym} is
  $B-\sigma(B)$ at $\Delta=0$, i.e.\ $y^N-(-x)^N$ by
  Eq.~\eqref{eq:Q0}, with the generating function $q-p=(x+y)tpq$. For
  $z=tpq/(2g)$ one finds $\delta z=-2g(x+y)z^2$, so that
  $(\delta+2g(x+y))\varphi(z)=2g(x+y)(\varphi-z^2\varphi')$ for any
  function $\varphi$, and the equation for the linear part becomes
  Euler's equation $z^2\varphi'=\varphi-z/2$, with the solution
  \begin{align}
    \sum_Nt^N D\big|_{\Delta^1}=\Delta\,\varphi(z),\qquad
    \varphi(z)=\frac12\sum_{k\ge1}(k-1)!\,z^k,\qquad z=\frac{t}{2g\,(1+xt)(1-yt)} .
    \label{eq:D1}
  \end{align}
  Expanding $z^k$ in binomial series gives, at depth $k=N-i-j\ge1$,
  \begin{align}
    D_{i,j}\big|_{\Delta^1}=(-1)^i\,\frac{(k-1)!}{2^{k+1}g^{k}}\binom{i+k-1}{k-1}\binom{j+k-1}{k-1}\,\Delta ,
    \label{eq:D1coef}
  \end{align}
  which reproduces the linear terms of Eq.~\eqref{eq:closedforms} and
  grows factorially with the depth.  For the linear part of $A$ one
  conjugates the first equation of \eqref{eq:s1234} with $P$, using
  $Ph_\pm P=h_\mp$. Then $PAP$ satisfies the equation of $D$ with the
  source $B-C$ replaced by $P(C-B)P$, which at $\Delta=0$ is
  $(a_-^\dagger)^N-(-a_+)^N$. The operator $\delta+2g(x+y)$ is
  invariant under $(x,y,g)\to(-x,-y,-g)$, and in terms of generating
  functions the new source is the old one after this substitution,
  with $t\to t/(1+2gt)$ and divided by $-(1+2gt)$, hence so is the
  solution.  Conjugating back with $P$, which replaces the symbols
  $x,y$ of $PAP$ by $2g-x$ and $-2g-y$, gives
  \begin{align}
    \sum_Nt^N A\big|_{\Delta^1}=-\frac{\Delta}{1+2gt}\,\varphi\bigl(-(1+2gt)\,z\bigr)
    =\frac{\Delta}{2}\sum_{k\ge1}(k-1)!\,(-1)^{k+1}(1+2gt)^{k-1}z^k .
    \label{eq:A1}
  \end{align}
  The part of $B$ of order $\Delta^2$ is then determined by
  $\mathcal{L}B|_{\Delta^2}=\frac\Delta2(A-D)|_{\Delta^1}$.  In the
  variables $u=x-2g$ (the symbol of $a_-^\dagger$), $y$ and $t$, the
  operator $\mathcal{L}$ acts on generating functions as
  $-(u\partial_u-y\partial_y+t\partial_t)$, which multiplies a
  monomial $u^ay^bt^N$ by $-(a-b+N)=-(2a+k)$, $k=N-a-b$, the pivot of
  Sec.~\ref{sec:recursion} up to the factor $2g$. It is inverted by
  $F\mapsto-\int_0^1\frac{d\lambda}{\lambda}F(\lambda
  u,y/\lambda,\lambda t)$. Carrying out the integral on
  Eqs.~\eqref{eq:D1} and \eqref{eq:A1} term by term gives, for
  $k=N-i-j\ge2$,
  \begin{align}
    B_{i,j}\big|_{\Delta^2}={}&(-1)^{i+1}\frac{\Delta^2}{16g^2}\sum_{n=1}^{k-1}(2g)^{k-2n}\,n!
                                \binom{i+n}{n}\binom{j+n}{n}\,\tau_{n,k}(i),
                                \nonumber\\
    \tau_{n,k}(i)={}&\sum_{r=0}^{k-n-1}\frac{(-1)^r}{r+1}\binom{n+i+r}{r}\binom{2i+k+r}{r+1}^{-1}
                      \Bigl[(-1)^n\binom{n}{k-n-1-r}-\delta_{r,k-n-1}\Bigr] .
                      \label{eq:B2}
  \end{align}
  For $k=2$ this is $(-1)^i\frac{\Delta^2}{16g^2}(j+1)$, Eq.~(27) of
  Ref.~\cite{xie_general_2022}. The denominators are products of the
  pivots $2i+k,\,2i+k+1,\dots$. The sum is nevertheless a polynomial
  in $i$ in all cases that we have examined, in accordance with
  Sec.~\ref{sec:depth}. Higher orders follow in the same way, or
  directly from the closed formula.

  \subsection{Explicit operators and data set}
  \label{sec:data}

  The entries $C$ and $A$ follow from $C=\sigma(B)$ and from the third
  equation of \eqref{eq:s1234},
  $A_{i,j}=D_{i,j}+\frac{2}{\Delta}[(j-i-N)B_{i,j}+2g(i+1)B_{i+1,j}]$,
  which produces no negative powers of $\Delta$ since
  $B_{i,j}=O(\Delta^2)$ for $(i,j)\ne(0,N)$. Written out in $\Delta$
  and $g$, the four entries for $N=8$ alone contain $328$ monomials,
  of which $111$ (those of $B$ and of the $D_{i,j}$ with $i\le j$) are
  independent, and we do not list the operators here. The complete
  operators $A$, $B$, $C$, $D$ for all $N\le20$ and the depth
  polynomials $f_k$, $h_k$ for $k\le14$ are available in
  machine-readable form with exact rational
  coefficients~\cite{kollar_symmetry_2026}, together with the program
  that generates them from the recursion and with programs that verify
  them and check the results of this paper in exact
  arithmetic. Appendix~A of Ref.~\cite{xie_general_2022} gives the
  operators for $N=5$, which agree with ours except for one term in
  the constant part of $A$. The printed term
  $-(\Delta^3+18\Delta)/(16g^3)$ should presumably read
  $-(\Delta^3+36\Delta)/(32g^3)$, as Eqs.~(18) and~(19) of that
  reference give.

  The operator identities of Sec.~\ref{sec:formula} and the formulas
  of this section were checked by computer in exact rational
  arithmetic, as identities in $\Delta$ and $g$, with the programs of
  Ref.~\cite{kollar_symmetry_2026} (the intertwining
  relation~\eqref{eq:intertwine} for all $N\le30$).  The results of
  Sec.~\ref{sec:su11} were checked in the same arithmetic for rational
  values of the parameters, in terms of the boson operators of the
  two-photon model for $N\le6$ and in representations of the discrete
  series with several Bargmann indices for $N\le4$.

  \section{Summary and outlook}
  \label{sec:summary}

  The hidden symmetry of the asymmetric quantum Rabi model at integer
  bias $N$ has an explicit generator. In the Bogoliubov frame of Xie
  and Chen, the symmetry operator is the parity of the symmetric model
  times the continuant of the $N$ first-order operators $\cK-c_m\sz$,
  $c_m=\frac{N+1}2-m$, with the spin-$\frac{N-1}2$ weights
  $m(N-m)/4$. It is a staircase in the bias that connects $H_N$ to its
  mirror image $H_{-N}$ and closes precisely at integer bias. At
  $\Delta=0$ it reduces to the $N$th powers of the two shifted ladder
  operators, which exchange the members of the degenerate pairs of the
  two displaced oscillators. Since the proof needs only the
  relations~\eqref{eq:relations}, the same continuant gives the
  symmetry operators of the asymmetric two-photon and two-mode quantum
  Rabi models for every order. The closed formula solves the recursion
  relations of Xie and Chen for the coefficients, which depend on $N$
  only through their distance from the highest degree.

  The closed formula raises several questions that are not treated
  here. The first concerns the polynomial $p_N$ in $J^2=p_N(H)$, which
  fixes the eigenvalues of $J$, its relation to the constraint
  polynomials of the Juddian solutions conjectured in
  Ref.~\cite{reyesbustos_degeneracy_2022}, and the kernel of
  $J$. Others concern the consequences of the formula for the
  spectrum, in particular for the $N$ lowest levels, which are
  unpaired at $\Delta=0$, and for matrix elements between the two
  symmetry sectors. These questions will be addressed in a separate
  publication. For the two-photon and two-mode models the
  corresponding polynomial is known only in the lowest cases, and its
  relation to the constraint
  polynomials~\cite{reyesbustos_twophoton_2026} is open. For the
  biased Dicke model and the anisotropic and Rabi-Stark models, where
  symmetry operators are known for small
  biases~\cite{lu_hidden_2021,lu_hidden_2022}, it remains to be seen
  whether relations of the form~\eqref{eq:relations} hold or which
  modification replaces them.

  \section*{Acknowledgments}

  The text and derivations in this paper, as well as the programs of
  Ref.~\cite{kollar_symmetry_2026} that verify them as described in
  Sec.~\ref{sec:data}, were developed in dialogue with Anthropic's
  Claude Fable~5.1 (November 2026) and Opus~5.5 (October 2026). The
  authors formulated the problems, directed and checked the work, and
  take full responsibility for the content.

  \section*{Data availability}

  The symmetry operators for $N\le20$ and the depth polynomials for
  $k\le14$ with exact rational coefficients, the program that
  generates them, and the programs that verify them and check the
  results of this paper are available at
  Ref.~\cite{kollar_symmetry_2026}.


\end{document}